\documentclass{article}
\usepackage[final]{graphicx}
\usepackage{amsmath,amssymb}

\begin{document}

\title{Quantum logic and weak values}
\author{Riuji Mochizuki\thanks{E-mail: rjmochi@tdc.ac.jp}\\
Laboratory of Physics, Tokyo Dental College,\\ Tokyo, Japan }

\setlength{\baselineskip}{20pt}

\maketitle

\begin{abstract}
\setlength{\baselineskip}{20pt}

In this study, we study weak values from a quantum-logical viewpoint.  In addition, we examine the validity of the counterfactual statements of Hardy's paradox, which are based on weak values, and we show that these statements have not been validated. It is also shown that strange weak values may only appear if they are not (conditional) probabilities.
\[
\]
\end{abstract}

\newpage

\section{Introduction}
Since Aharonov et al.\cite{Aha2}\cite{Aha25}\cite{Aha3} developed the concepts of weak measurement and weak values, these ideas have attracted great attention.  In weak measurement, which differs from conventional  von Neumann-type measurement\cite{von}, the interaction between an observed system and a probe is considered to have no effect on the observed system when its weak coupling limit is taken.   Some authors\cite{Hosoya1}\cite{Tresser1} have even claimed that noncommuting observables could be measured simultaneously by weak measurement, and relations to Bell's inequality\cite{Bell} have also been discussed.  

 It is believed that weak measurement enables us to select both an initial state and a final state.  The weak value of an operator $\hat A$ for an initial state $|\Phi\rangle$ and a final state $|\Psi\rangle$ is defined as
\begin{equation}
\langle\hat A\rangle_{\Psi,\Phi}\equiv{\langle\Psi |\hat A|\Phi\rangle\over\langle\Psi |\Phi\rangle}.\label{eq:teigi}
\end{equation}
Recently, weak values have attracted attention due to both the values obtained by  weak measurement and their inherent physical meaning\cite{Hosoya2}.   For example, the counterfactual statements of Hardy's paradox\cite{Hardy} were interpreted with the help of weak values\cite{Aha4}, which were experimentally verified\cite{Irvine}\cite{Lundeen}\cite{Yokota} to agree with the values obtained by  the corresponding weak measurements.    Moreover,  strange weak values have been discussed by many authors\cite{Leggett}\cite{Duck}\cite{Hosoya2}\cite{Dressel}, but the conditions in which they appear have not been clarified.

In this study,  we study weak values from a quantum-logical viewpoint to clarify what they represent.  Moreover, we investigate the validity of the counterfactual statements of Hardy's paradox, which are based on weak values.  The following conclusions are reached:  (i) the weak value (\ref{eq:teigi}) that is associated with a projection operator $\hat A$ is a (conditional) probability if and only if $[|\Psi\rangle\langle \Psi |,\hat A]=0$,  $[|\Phi\rangle\langle \Phi |,\hat A]=0$ or $[|\Psi\rangle\langle\Psi |,|\Phi\rangle\langle\Phi |]=0$.  (ii) Strange weak values (complex numbers and real numbers that are not between 0 and 1) may only appear if they are not (conditional) probabilities.  (iii) The validity of the counter factual statements of Hardy's paradox has not been evaluated in terms of weak values.   

In the second section, we investigate the weak values of projection operators and present the conditions where they are regarded as conditional probabilities.  Hardy's paradox is examined in the third section, and its counterfactual interpretation based on weak values is also discussed there.  The last section is our conclusion.

\section{What are weak values?}

We examine the expectation value $\langle\Phi |\hat A |\Phi\rangle$ of an observable $\hat A$ for a state vector $|\Phi \rangle$.  Let $|\psi_j\rangle$ be the eigenvectors that correspond to the respective eigenvalues $\psi_j,\ j=1,2,\cdots$ of an observable $\hat \Psi$.   By assuming that  $1=\sum_j|\psi_j\rangle\langle\psi_j|$ and that $\langle\psi_j|\Phi\rangle\ne0$,  
\begin{equation}
\begin{split}
\langle\Phi |\hat A |\Phi\rangle &=\sum_j\langle\Phi |\psi_j\rangle\langle\psi_j |\hat A |\Phi\rangle \\
&=\sum_j|\langle\Phi |\psi_j\rangle|^2{\langle\psi_j |\hat A |\Phi\rangle\over\langle\psi_j |\Phi\rangle} \\
&=\sum_j{\rm Pr}(\psi_j|\Phi)\langle\hat A\rangle_{\psi_j,\Phi},\label{eq:0}
\end{split}
\end{equation}
where
\[
{\rm Pr}(\psi_j|\Phi)=|\langle\Phi |\psi_j\rangle |^2
\]
is the probability that the state $|\psi_j\rangle$ is found in the state $|\Phi\rangle$.  Thus, we can interpret the expectation value $\langle\Phi |\hat A |\Phi\rangle$ as a statistical average of the weak values $\langle\hat A\rangle_{\Psi_j,\Phi}$, and as a result, weak values are treated by many authors as the expectation values of $\hat A$ between the initial state $|\Phi\rangle$ and the final states $|\psi_j\rangle,\ j=1,2,\cdots$.   However, as shown below, we should not decide based exclusively on (\ref{eq:0})  whether weak values can be interpreted as probabilities or expectation values.

We write the proposition `an eigenvalue $a_i$ is obtained when an observable $\hat A$ is measured' as $A(a_i)$, and its corresponding projection operator is denoted $\hat A_i=|a_i\rangle\langle a_i|$.  Similarly, we define a proposition $\Psi(\psi_j)$ and a projection operator $\hat\Psi_j=|\psi_j\rangle\langle\psi_j|$.  A set of such propositions constitutes a $\sigma$-complete orthomodular lattice\cite{Svozil}\cite{Maeda}, as does the corresponding set of such projection operators.

Let $\hat A$ in (\ref{eq:0}) be the projection operator $ \hat A_i=|a_i\rangle\langle a_i |$.  Then,
\begin{equation}
\langle\Phi |\psi_j\rangle \langle \psi_j |a_i\rangle\langle a_i |\Phi\rangle ={\rm Pr}(\psi_j|\Phi )\langle\hat A_i\rangle_{\psi_j,\Phi}.\label{eq:2}
\end{equation}
A necessary and sufficient condition for the operator $\hat \Psi_j\hat A_i$ to be a projection operator is $[\hat \Psi_j,\hat A_i]=0$.  If and only if this condition is satisfied, $\hat \Psi_j\hat A_i$ corresponds to a proposition $\Psi(\psi_j)\wedge A(a_i)$ and the left-hand side of (\ref{eq:2}) is its probability for $|\Phi\rangle$, i.e., the probability of finding $|\psi_j\rangle$ {\it and} $|a_i\rangle$ in $|\Phi\rangle$.  Thus, the weak value $\langle\hat A\rangle_{\psi_j,\Phi}$ is the conditional probability of finding $|a_i\rangle$ in $|\Phi\rangle$ when $|\psi_j\rangle$ is found in $|\Phi\rangle$ if and only if $[\hat \Psi_j,\hat A_i]=0$.  Then,
\[
0\le\langle\Phi |\psi_j\rangle \langle \psi_j |a_i\rangle\langle a_i |\Phi\rangle\le{\rm Pr}(\psi_j|\Phi )\le 1,
\]
and hence,
\begin{equation}
0\le\langle\hat A_i\rangle_{\psi_j,\Phi}\le 1.
\end{equation}
As shown later, the weak values are actually 0 or 1 in such a case.  
We can interchange $|\psi_j\rangle$ and $|\Phi\rangle$ in the above discussion.  If $|\Phi\rangle\langle\Phi |$ and $\hat\Psi_j$ commute,  $\langle\hat A\rangle_{\psi_j\Phi}$ is the probability of finding $|a_i\rangle$ in $|\Phi\rangle$ (or in $|\psi_j\rangle$).  

If $[\hat \Psi_j,\hat A]\ne 0$, the projection operator that corresponds to a proposition $\Psi(\psi_j)\wedge A(a_i)$ is $\lim_{n\rightarrow \infty}(\hat\Psi_j \hat A_i)^n$\cite{Svozil}.  Instead, if we construct (for example) a hermitian operator $\hat{H}\equiv\hat A\hat\Psi\hat A$ and a projection operator $\hat H_k\equiv |h_k\rangle\langle h_k|$ where $\hat H |h_k\rangle =h_k |h_k\rangle$, then the proposition corresponding to $\hat H_k$ exists.  Nevertheless, this proposition is not expressed with the help of the $\Psi (\psi_j)$s and/or $A(a_i)$s.  In contrast, either $\hat \Psi_j\hat A_i$ is not a projection operator or it does not correspond to any propositions.  Thus, if any two of $\hat\Psi_j$, $\hat A_i$ and $|\Phi\rangle\langle\Phi |$ do not commute, we cannot interpret the left-hand side of (\ref{eq:2}) as a probability or the right-hand side of (\ref{eq:0}) as a sum of probabilities.   Therefore, in such cases,  $\langle\hat A_i\rangle_{\psi_j,\Phi}$ is not the conditional probability of finding $|a_i\rangle$ in $|\Phi\rangle$ when $|\psi_j\rangle$ is found in $|\Phi\rangle$,

We comment on strange weak values.  We divide $\langle\Phi |\hat \Psi_j\hat A_i|\Phi \rangle$ into its real part and imaginary part as follows:
\begin{equation}\begin{split}
\langle\Phi |\hat \Psi_j\hat A_i|\Phi \rangle=&\langle\Phi |{1\over 2}(\hat \Psi_j\hat A_i+ \hat A_i\hat \Psi_j)|\Phi \rangle \\
&+\langle\Phi |{1\over 2} [\hat\Psi_j ,\hat A_i] |\Phi\rangle.
\end{split}
\end{equation}
Thus, the weak value
\begin{equation}
\langle \hat A\rangle_{\psi_j,\Phi}={\langle\Phi |\hat\Psi_j\hat A_i |\Phi\rangle\over{\rm Pr}(\psi_j|\Phi)}\label{eq:wv}
\end{equation}
is a complex number if $\langle\Phi |[\hat \Psi_j,\hat A_i]|\Phi\rangle\ne 0$.  However, this value becomes real if $\langle\Phi |[\hat \Psi_j,\hat A_i]|\Phi\rangle= 0$.  Here, we should pay attention to the fact that even if $\langle |[\hat \Psi_j,\hat A_i]|\rangle= 0$ for some states, this is not a sufficient condition for $[\hat \Psi_j,\hat A_i]=0$, i.e., this condition does not ensure that  
$\hat \Psi_j\hat A_i$ is a projection operator and possesses the corresponding proposition.   If  $\langle\Phi |[\hat \Psi_j,\hat A_i]|\Phi\rangle= 0$ and any pair of $\hat\Psi_j$, $\hat A_i$ and $|\Phi\rangle\langle\Phi |$ do not commute, (\ref{eq:wv}) may be more than 1 or less than 0.  This possibility is not strange because (\ref{eq:wv}) is not a (conditional) probability as shown above.  We will encounter such a situation in the next section.  

To corroborate the above conclusion, we reexamine $\langle\Phi |\hat A|\Phi\rangle$.  When $\hat A=\hat A_i$,
\begin{equation}
\begin{split}
\langle\Phi |\hat A_i|\Phi\rangle&=\langle\Phi |\hat A_i\hat A_i|\Phi\rangle\\
&=\sum_j\langle\Phi |\hat A_i\hat \Psi_j\hat A_i|\Phi\rangle\\
&=\sum_j{\rm Pr}(\psi_j|\Phi)|\langle\hat A_i\rangle_{\psi_j,\Phi}|^2.\label{eq:0dash}
\end{split}\end{equation}
$\langle\Phi|\hat A_i\hat\Psi_j\hat A_i|\Phi\rangle =\langle\Phi|\hat \Psi_j\hat A_i|\Phi\rangle$ if  $\hat A_i$ and $\hat \Psi_j$ commute.  Then, by comparing (\ref{eq:0dash}) and (\ref{eq:0}), it is clear that $\langle\hat A_i\rangle_{\psi_j,\Phi}=0$ or $1$.  Conversely, if $\langle\hat A_i\rangle_{\psi_j,\Phi}\ne |\langle\hat A_i\rangle_{\psi_j,\Phi} |^2$, it is obvious that at least one of the following two statements is false: `$\langle\hat A_i\rangle_{\psi_j,\Phi}$ is the expectation value of $\hat A_i$ between an initial state $|\Phi\rangle$ and a final state $|\psi_j\rangle$'; `$|\langle\hat A_i\rangle_{\psi_j,\Phi}|^2$ is the expectation value of $\hat A_i$ between an initial state $|\Phi\rangle$ and a final state $|\psi_j\rangle$'.  We have shown above that the former statement is false if the operators do not commute, and we will show below that the latter statement is also false if they do not commute.

The above discussion can be straightforwardly applied to other observables, such as $\hat A=\sum_ia_i\hat A_i$.   Thus, it is obvious that if $\hat \Psi_j$ and $\hat A$ do not commute, then the weak value $\langle\hat A\rangle_{\psi_j,\Phi}$ is not the conditional expectation value of $\hat A$ for $|\Phi\rangle$ when $|\psi_j\rangle$ is found in $|\Phi\rangle$.  We then must ask what the weak values are.  As written by Aharonov et al.\cite{Aha1},
\begin{equation}
|\langle\hat A_i\rangle_{\psi_j,\Phi} |^2 = {{\rm Pr}(a_i|\psi_j){\rm Pr}(a_i|\Phi)\over {\rm Pr}(\psi_j|\Phi)}.\label{eq:jijou}
\end{equation}
Because the denominator of the right-hand side does not depend on $a_i$, $|\langle\hat A_i\rangle_{\psi_j,\Phi} |^2$ gives the product of two independent probabilities {\rm Pr}($a_i |\psi_j$) and {\rm Pr}($a_i |\Phi$) (divided by Pr$(\psi_j|\Phi )$).  It is worth noting that (\ref{eq:jijou}) is not a conditional probability if $[\hat \Psi_j,\hat A_i]\ne0$.  To see this fact, we rewrite (\ref{eq:jijou}) as
 \begin{equation}
|\langle\hat A_i\rangle_{\psi_j,\Phi} |^2 = {\langle\Phi |\hat A_i\hat \Psi_j\hat A_i |\Phi\rangle \over {\rm Pr}(\psi_j|\Phi)}.\label{eq:jijou2}
\end{equation}
The right-hand side of this equation is the expectation value of {\it one} observable $\hat A_i\hat\Psi_j\hat A_i$ divided by Pr$(\psi_j|\Phi )$.  If $[\hat \Psi_j,\hat A_i]\ne0$, then $\hat A_i\hat\Psi_j\hat A_i$ corresponds to no proposition, and consequently,  (\ref{eq:jijou}) is not a conditional probability because $\hat A_i\hat\Psi_j\hat A_i$ is not a projection operator.  
More generally, 
\begin{equation}
|\langle\hat A\rangle_{\psi_j,\Phi} |^2 = {\langle\Phi |\hat A\hat \Psi_j\hat A |\Phi\rangle \over {\rm Pr}(\psi_j|\Phi)},\label{eq:jijou3}
\end{equation}
though the quantity that corresponds to a hermitian operator $\hat A\hat\Psi_j\hat A$ is not known.

Before applying the discussion in this section to Hardy's paradox, we comment on the commutativity of operators in experiments.  Because error is not avoidable there, {\it the commuting operators} and {\it the noncommuting operators} should be continuously connected.  If two projection operators $\hat X_i\equiv |x_i\rangle\langle x_i|$ and $\hat Y_j\equiv |y_j\rangle\langle y_j|$ have a very small commutator $[\hat Y_j,\hat X_i]$,  we can regard them as {\it commuting} and $\hat Y_j\hat X_i$ has a corresponding proposition.  More rigorously, because of the identity
\begin{equation}
\hat Y_j\hat X_i\hat Y_j\hat X_i=(1-\langle [\hat Y_j, \hat X_i]\rangle_{y_jx_i})\hat Y_j\hat X_i,
\end{equation}
$\hat Y_j\hat X_i$ can be regarded as a projection operator and $\langle\Xi |\hat Y_j\hat X_i|\Xi\rangle$ can be considered as the probability of a proposition $Y(y_j)\wedge X(x_i)$ for a state $|\Xi\rangle$ if $|\langle [\hat Y_j, \hat X_i]\rangle_{y_jx_i}|$ is smaller than its relative error.  For  $\hat X\equiv\sum_ix_i\hat X_i$,  $\langle\Xi |\hat Y_j\hat X|\Xi\rangle$ can be interpreted as an expectation value if $\sum_i\langle\Xi |\hat Y_j\hat X_i|\Xi\rangle x_i\langle [\hat Y_j, \hat X_i]\rangle_{y_jx_i}$ is smaller than its absolute error.

\section{Hardy's paradox}

Recently, the counter factualstatements of Hardy's paradox were interpreted with the help of weak values\cite{Aha4}, and it was ascertained that they agreed with the values obtained by the corresponding weak measurement\cite{Irvine}\cite{Lundeen}\cite{Yokota}. However, this agreement does not warrant the validity of the interpretation, as the meaning of the weak measurements has been interpreted only operationally.    Thus, we should not explain weak values based on the corresponding weak measurement.  Rather, the meaning of a weak measurement should be clarified by investigating the corresponding weak values.

 We investigate the weak values in Hardy's paradox based on the discussion in the previous section.  

\includegraphics{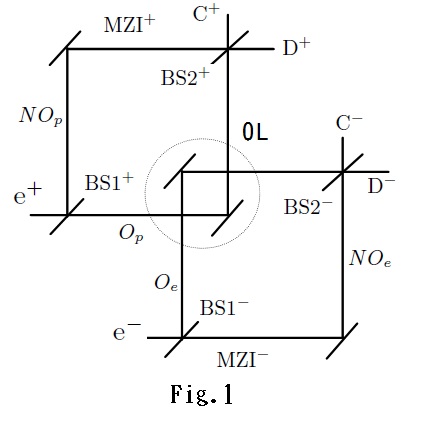}
As shown in Fig. 1, a device composed of an electron Mach-Zehnder interferrometer (MZI$^-$) and a similar machine with positrons (MZI$^+$) is examined.  OL is the domain where these two MZIs overlap.  We assume that pair annihilation must occur if an electron (e$^-$) and a positron (e$^+$) exist simultaneously in OL.  The length between BS1$^{-(+)}$ and BS2$^{-(+)}$ is adjusted to let e$^-$ (e$^+$) be detected by a detector C$^{-(+)}$ without exception in a solo MZI$^{-(+)}$ experiment.  Conversely, detection by a detector D$^{-(+)}$ implies that  obstacles exist on either path.

We consider the case where the pair annihilation does not occur and e$^{-}$ and e$^+$ are detected by D$^-$ and D$^+$, respectively.  The initial state $|\Phi\rangle$ and the final state $|\Psi\rangle$ are  defined as
\begin{equation}\begin{split}
|\Phi\rangle = {1\over\sqrt{3}}\Big[ &|O_p,NO_e\rangle + |NO_p,O_e\rangle\\
& + |NO_p,NO_e\rangle\Big],
\end{split}
\end{equation}
\begin{equation}\begin{split}
|\Psi\rangle =& {1\over2}\Big[ |O_p,O_e\rangle - |O_p,NO_e\rangle\\
& - |NO_p,O_e\rangle + |NO_p,NO_e\rangle\Big],
\end{split}
\end{equation}
where $O$ and $NO$ are abbreviations of `{\it Through OL}' and `{\it Not through OL}', respectively.  
 Then $\big|\langle\Psi |\Phi\rangle\big|^2  = {1\over 12}$ by ordinary quantum mechanical calculation. However, the weak values are
\begin{equation}
\langle \hat N^{+,-}_{O,O}\rangle_{\Psi, \Phi} = 0,\label{eq:oo}
\end{equation}
\begin{equation}
\langle \hat N^{+,-}_{O,NO}\rangle_{\Psi, \Phi} =\langle \hat N^{+,-}_{NO,O}\rangle_{\Psi, \Phi} = 1,
\end{equation}
\begin{equation}
\langle \hat N^{+,-}_{NO,NO}\rangle_{\Psi, \Phi} = -1,\label{eq:nono}
\end{equation}
\begin{equation}
\langle \hat N^{\pm}_O\rangle_{\Psi, \Phi} = 1,\label{eq:ohitotsu}
\end{equation}
\begin{equation}
\langle \hat N^{\pm}_{NO}\rangle_{\Psi, \Phi} = 0,\label{eq:nohitotsu}
\end{equation}
where
\begin{equation}\begin{split}
\hat N^{+,-}_{O(NO),O(NO)}=|&O(NO)_p,O(NO)_e\rangle\\
\otimes&\langle O(NO)_p,O(NO)_e|,\label{eq:number2}\end{split}
\end{equation}
\begin{equation}
\hat N^{+}_{O(NO)}=\hat N^{+,-}_{O(NO),O}+\hat N^{+,-}_{O(NO),NO}, \label{eq:number1}
\end{equation}
\begin{equation}
\hat N^{-}_{O(NO)}=\hat N^{+,-}_{O,O(NO)}+\hat N^{+,-}_{NO,O(NO)}. \label{eq:number3}
\end{equation}

It is easily verified that any two of $\hat \Psi\equiv|\Psi\rangle\langle\Psi |$, $\hat\Phi\equiv|\Phi\rangle\langle\Phi |$ and any one of the operators defined in (\ref{eq:number2}) $\sim$ (\ref{eq:number3}) do not commute.  For example, though $\langle\Phi|[\hat \Psi ,\hat N^{+,-}_{NO,NO}]|\Phi\rangle=\langle\Psi[\hat \Phi ,\hat N^{+,-}_{NO,NO}]|\Psi\rangle =0$,
\begin{equation}\begin{split}
\hat\Psi\hat N^{+,-}_{NO,NO}\hat \Psi\hat N^{+,-}_{NO,NO}&={1\over 4}\hat \Psi\hat N^{+,-}_{NO,NO},\\
\hat\Phi\hat N^{+,-}_{NO,NO}\hat \Phi\hat N^{+,-}_{NO,NO}&={1\over 3}\hat \Phi\hat N^{+,-}_{NO,NO}.\\ 
\label{eq:zero}\end{split}
\end{equation}
Therefore, $\langle \hat N^{+,-}_{NO,NO}\rangle_{\Psi, \Phi}$ cannot be regarded as the  conditional probability of finding both e$^-$ and e$^+$ on {\it NO}s between the initial state $|\Phi\rangle$ and the final state $|\Psi\rangle$. The discussions of $\hat N^{+,-}_{O,NO}$, $\hat N^{+,-}_{NO,O}$ and $\hat N^{+,-}_{O,O}$ are roughly equivalent.  Thus, regardless of which operator is used, even with the help of the weak values (\ref{eq:oo}) $\sim$ (\ref{eq:nohitotsu}) we have no right to evaluate the validity of the counterfactual statement ` e$^-$ must pass through OL to ensure that e$^+$ is detected by D$^+$ and vice versa.  Nevertheless, both e$^-$ and e$^+$ cannot simultaneously pass through OL  because they must be annihilated together if they encounter each other'.  All we can verify with the help of the weak values (\ref{eq:oo}) $\sim$ (\ref{eq:nohitotsu}) is that
\begin{equation}\begin{split}
&{\rm Pr}(N^{+,-}_{O,O}|\Psi){\rm Pr}(N^{+,-}_{O,O}|\Phi)\\
&:{\rm Pr}(N^{+,-}_{O,NO}|\Psi){\rm Pr}(N^{+,-}_{O,NO}|\Phi)\\
&:{\rm Pr}(N^{+,-}_{NO,O}|\Psi){\rm Pr}(N^{+,-}_{NO,O}|\Phi)\\
&:{\rm Pr}(N^{+,-}_{NO,NO}|\Psi){\rm Pr}(N^{+,-}_{NO,NO}|\Phi)\\
&=0:1:1:1,\label{eq:0111}
\end{split}
\end{equation}
based on (\ref{eq:jijou}).
 We can interpret the weak values in the Three-box paradox\cite{Aha6}\cite{Vaidman}\cite{Resch} similarly.

\section{Conclusion}
In this study, we studied weak values from a quantum-logical viewpoint.  In addition, we examined the validity of the counterfactual interpretations of Hardy's paradox not operationally but in an investigation of the corresponding weak values.    We then concluded that we were not able to evaluate the interpretations of Hardy's paradox even with the help of  weak values because they are not conditional probabilities.  In general, a weak value $\langle\hat A\rangle_{\Psi.\Phi}$ is not a (conditional) probability or a (conditional) expectation value if any two of $\hat A$, $|\Psi\rangle\langle\Psi |$ and $|\Phi\rangle\langle\Phi |$ do not commute.  




\begin{thebibliography}{99}
\bibitem{Aha2}Y. Aharonov, D. Z. Albert and L. Vaidman, {\it Phys. Rev. Lett.} {\bf 60}, 1351 (1988)
\bibitem{Aha25}Y. Aharonov and L. Vaidman, {\it Phys. Rev. A} {\bf 41}, 11 (1990)
\bibitem{Aha3}Y. Aharonov and A. Botero, {\it Phys. Rev. A} {\bf 72}, 052111 (2005)
\bibitem{von}J. von Neumann, ``Die Mathematische Grundlagen der Quantenmechanik," Springer Verlag, Verlin, 1932
\bibitem{Hosoya1}A. Hosoya and M. Koga, {\it J. Phys. A} {\bf 44}, 415303 (2011)
\bibitem{Tresser1}C Tresser, {\it Eur. Phys. J.} {\bf D58}, 385 (2010)
\bibitem{Bell}J. S. Bell, {\it Physics} {\bf 1}, 195 (1964)
\bibitem{Hosoya2}A. Hosoya and Y. Shikano, {\it J. Phys. A: Math. Theor.} {\bf 43}, 385307 (2010)
\bibitem{Hardy}L. Hardy, {\it Phys. Rev. Lett.} {\bf 68}, 2981 (1992)
\bibitem{Aha4}Y. Aharonov, A. Botero, S. Popescu, B. Reznik and J. Tollaksen, {\it Phys. Lett.} {\bf 301}, 130 (2002)
\bibitem{Irvine}W. T. M. Irvine, J. F. Hodelin, C. Simon and D. Bouwmeester, {\it Phys. Rev. Lett.} {\bf 95}, 030401 (2005)
\bibitem{Lundeen}J. S. Lundeen and A. M. Steinberg, {\it Phys. Rev. Lett.} {\bf 102}, 020404 (2009)
\bibitem{Yokota}K. Yokota, T, Yamamoto, M. Koashi and N. Imoto, {\it New J. Phys.} {\bf 11}, 033011 (2009)
\bibitem{Leggett}A. J. Leggett, {\it Phys. Rev. Lett.} {\bf 62}, 2325 (1989)
\bibitem{Duck}I. M. Duck, P. M. Stevenson and E. C. G. Sudarshan, {\it Phys. Rev. D} {\bf 40}, 2112 (1989)
\bibitem{Dressel}J. Dressel and A. N. Jordan, {\it Phys. Rev. A} {\bf 85}, 012107 (2012)
\bibitem{Svozil}K. Svozil, ``Quantum Logic," Springer-Verlag, Singapore City (1998)
\bibitem{Maeda}S. Maeda, ``Lattice Theory and Quantum Logic," Maki-Shoten, Tokyo (1980)
\bibitem{Aha1}Y. Aharonov, P. G. Bergmann and J. L. Lebowitz, {\it Phys. Rev. }{\bf 134}, B1430 (1964)
\bibitem{Aha6}Y. Aharonov and L. Vaidman, {\it J. Phys. A: Math. Gen.} {\bf 24}, 2315 (1991)
\bibitem{Vaidman}L. Vaidman, {\it Found. Phys.} {\bf 26}, 895 (1996)
\bibitem{Resch}K. J. Resch, J. S. Lundeen and A. M. Steinberg, {\it Phys. Lett. A} {\bf 324}, 125 (2004)




\end{thebibliography}
\end{document}